\newif\iffigs\figstrue

\documentstyle[12pt]{article}
\def\semi{;\ }
\def\ppnumber{\vbox{\baselineskip14pt
\hbox{DUKE-TH-96-107}\hbox{HUTP-96/A012}\hbox{hep-th/9603161}}}
\def\ppdate{\vbox{\hbox{March, 1996}\hbox{(Revised 5/96)}}}
\def\pplogo{\vbox{\kern-\headheight\kern -35pt
\halign{##&##\hfil\cr&{\ppnumber}\cr\rule{0pt}{2.5ex}&\ppdate\cr}
}}

\makeatletter

\@addtoreset{equation}{section}
\def\eqref#1{(\ref{#1})}

\date{}
\def\ps@firstpage{\ps@empty \def\@oddhead{\hss\pplogo}%
  \let\@evenhead\@oddhead 
}
\def\maketitle{\par
 \begingroup
 \def\thefootnote{\fnsymbol{footnote}}
 \def\@makefnmark{\hbox
 to 0pt{$^{\@thefnmark}$\hss}}
 \if@twocolumn
 \twocolumn[\@maketitle]
 \else \newpage
 \global\@topnum\z@ \@maketitle \fi\thispagestyle{firstpage}\@thanks
 \endgroup
 \setcounter{footnote}{0}
 \let\maketitle\relax
 \let\@maketitle\relax
 \gdef\@thanks{}\gdef\@author{}\gdef\@title{}\let\thanks\relax}

\def\abstract{\if@twocolumn
\section*{Abstract}
\else \small
\begin{center}
{\bf ABSTRACT}
\end{center}
\quotation
\fi}

\def\thebibliography#1{\section*{References\@mkboth
 {REFERENCES}{REFERENCES}}\small\list
 {[\arabic{enumi}]}{\settowidth\labelwidth{[#1]}\leftmargin\labelwidth
 \advance\leftmargin\labelsep
 \usecounter{enumi}}
 \def\newblock{\hskip .11em plus .33em minus .07em}
 \sloppy\clubpenalty4000\widowpenalty4000
 \sfcode`\.=1000\relax}

\makeatother

\begin{document}
\iffigs
  \input epsf
\else
  \message{No figures will be included. See TeX file for more
information.}
\fi

\def\ord{\mathop{\rm ord}\nolimits}
\def\Q{{\bf Q}}
\def\Z{{\bf Z}}

\setcounter{page}0
\title{\LARGE Compactifications of F-Theory\\
on Calabi--Yau Threefolds -- II\\[10mm]}
\author{
\normalsize David R. Morrison\\[0.35cm]
\normalsize \it Department of Mathematics, Duke University\\
\normalsize \it Durham, NC 27708, USA\\[0.35cm]
\normalsize and\\[0.35cm]
\normalsize Cumrun Vafa\\[0.35cm]
\normalsize \it Lyman Laboratory of Physics, Harvard University\\
\normalsize \it Cambridge, MA 02138, USA\\[10mm]
}

\maketitle

\def\Large{\large}
\def\LARGE{\large\bf}

\begin{abstract}

We continue our study of compactifications of F-theory on Calabi--Yau
threefolds.  We gain more insight into F-theory duals of heterotic strings
and provide a recipe for building F-theory duals for arbitrary heterotic
compactifications on elliptically fibered manifolds.   As a byproduct we
find that string/string duality in six dimensions gets mapped to fiber/base
exchange in F-theory.  We also construct a number of new $N=1$, $d=6$
examples of F-theory vacua and study transitions among them.  We find that
some of these transition points correspond upon further compactification to
4 dimensions to transitions through analogues of Argyres--Douglas points of
$N=2$ moduli.  A key idea in these transitions is the notion of classifying
$(0,4)$ fivebranes of heterotic strings.

\end{abstract}

\vfil\break

\section{Introduction}

Compactifying
F-theory on elliptic Calabi--Yau threefolds leads to theories in six
dimensions with $N=1$ supersymmetry \cite{vft}.  In this paper we continue
our studies of such compactifications
extending our previous results
\cite{mv} by considering a more general class of Calabi--Yau threefolds.

The case considered in \cite{mv} was motivated by considering
duals to perturbative heterotic string vacua compactified
on $K3$.   In particular
for $E_8\times E_8$ compactifications
we found that for instanton numbers $(12-n,12+n)$ for the $E_8$'s,
the corresponding F-theory dual
is compactification on the unique\footnote{This assertion of uniqueness
assumes that the elliptic fibration has a section; furthermore, we
treat F-theory models which differ by conifold transitions as being
the same, as discussed in section 3 below.}
elliptic Calabi--Yau threefold with the base
being the rational ruled surface ${\bf F}_n$.   Moreover the $SO(32)$
heterotic string was conjectured to be dual to the $E_8\times E_8$
model with
instanton numbers $(8,16)$.

In this paper we continue our studies in two directions:  First of
all we gain a better insight into the simple correspondence
we found between instanton numbers and the degree of the rational
ruled surface.  In this way we find a recipe for how to {\it build up}\/
F-theory duals to heterotic vacua in arbitrary dimensions
compactified on elliptic manifolds
with given bundle data.  This investigation also leads us
to reinterpret the F-theory/heterotic duality in 8 dimensions \cite{vft}
as the statement of exchange symmetry between the fiber and the base
of F-theory thus taking a step in geometrizing string/string duality.

We then consider other vacua of F-theory.  Among these one can obtain all
the non-perturbative heterotic vacua with fivebranes
turned on \cite{dmw,sen,sw}
and some additional ones which do not
have any heterotic analogue.  Among these we find theories which have
no tensor multiplets at all (which has
also been noted in \cite{sw}).  Also allowing for mild orbifold
singularities in the base manifold we find that there
are interesting classes of models which are stuck in phases
with tensionless strings (and  form an isolated class
of $N=1$ vacua in $d=6$).
In this way we find that the F-theory
is a unifying setup for all the $N=1$ $d=6$ vacua.

We then talk about various types of transitions among these vacua.
It has been suggested in \cite{sw,han}
that transitions among such vacua are
through regions with tensionless strings.  Since transitions among Calabi--Yau
threefolds
are well studied, and given that F-theory compactifications
are on elliptic Calabi--Yau threefolds one can examine such transitions
from the 4 dimensional viewpoint.  In this way we find that upon
compactifications to 4 dimensions the relevant six dimensional
transitions get mapped to transitions of one of two types, one of
which passes through an
Argyres--Douglas point in the moduli of $N=2$ theories.
Even though the physics of such transitions are not
yet clear, the mathematical aspect of the extremal transition
are fairly well understood.  Moreover we find that at least a class
of such transitions are related to exceptional groups $E_n$
(extending $n$ all the way down from 8 to 1).  In particular
we find a strong hint that in the heterotic theory there are
different types of (0,4) 5-branes.  Roughly speaking they should be in
one to one correspondence with the degenerating types of $(0,4)$
conformal theories which are the perturbative heterotic vacua.
We tentatively identify the $E_n$ type degeneration as the F-theory dual
of heterotic string fivebranes corresponding to zero size instanton of gauge
group $E_n$.  We find examples of all such transitions.
The most relevant
one for the transition  among $E_8\times
E_8$ perturbative heterotic vacua  are the $E_8$ type.  Clearly
there are other types of $(0,4)$ fivebranes and we believe
we have taken a step in classifying
them and their F-theory duals.  We believe that further studies in extending
this classification are very important.

The organization of this paper is as follows:  In section 2 we review
F-theory/heterotic duality in 8 dimensions.  In section 3 we show
how the $N=1$ F-theory/heterotic duality in 6 dimensions can be
deduced from the 8-dimensional duality.  In section 4 we make some
general remarks about the spectrum of F-theory compactifications
on Calabi--Yau threefolds.  In section 5 we discuss the conditions
imposed on the bases of the elliptic threefold by the condition
of vanishing $c_1$ of the Calabi--Yau and find that in the case
of the smooth bases they are all classified and are given by
rational surfaces.  In section 6 we give a number of examples.
These include, apart from the ${\bf F}_n$ already discussed in section
3, ${\bf P}^2$ blown up at a number of points, Voisin--Borcea examples,
as well as orbifold examples.  Finally in section 7 we discuss
singularities in the moduli space and transitions among the vacua.

\section{F-Theory/Heterotic Duality in 8 Dimensions}

In this section we recall the F-theory/heterotic duality in 8-dimensions
\cite{vft}.  We consider heterotic strings compactified
to 8 dimensions on $T^2$.  The moduli of this theory are characterized
by specifying the complex structure $\tau$ and complexified K\"ahler
structure $\rho$ of $T^2$ and in addition
Wilson lines of $E_8\times E_8$ (or $SO(32)$) on
$T^2$ characterized by 16 complex parameters.  Altogether
we thus have 18 complex parameters.   They are given globally
by the moduli space \cite{narain,narainetal}
\begin{equation}\label{globalmod}
SO(18,2;{\bf Z})\backslash SO(18,2)/SO(18)\times SO(2).
\end{equation}
In addition there is a coupling constant of heterotic strings
parametrized by a positive real number $\lambda^2\in {\bf R}^+$.  The
dual on the F-theory
side is given by $K3$ compactifications which admit an elliptic fibration.
The elliptic fiber, a torus, can be represented by
$$y^2=x^3+ax +b$$
which expresses it
as a double cover of the complex plane $x$.  The $K3$ surface can then
be viewed
as fiber space where the base is ${\bf P}^1$ and the fiber is the
torus. Let us denote the coordinate of the ${\bf P}^1$ by $z$. Then
the elliptic fibration is specified by
$$y^2=x^3+f(z) x+g(z)$$
where $f$ is of degree $\le8$ and $g$ is of degree $\le12$.
Altogether we have 9+13 parameters specifying the polynomials $f$ and $g$,
but we must mod out by the equivalence given by the $SL(2,{\bf C})$ action
on ${\bf P}^1$ (3 parameters) and by one rescaling ($y\rightarrow \mu^3 y,
x\rightarrow \mu^2 x, f\rightarrow \mu^4 f,
g\rightarrow\mu ^6g$ for $\mu\in{\bf C}^*$),
so that we end up with 18 complex parameters.
It is well-known
that this is also described\footnote{What is well-known is that this
is described by some space of the form
$\Gamma\backslash SO(18,2)/SO(18)\times SO(2)$; it would be interesting
to verify that $\Gamma=SO(18,2;{\bf Z})$.}
 by the above moduli space.
Moreover there is an additional real parameter which is the size $A$ of ${\bf
P}^1$ and we identify the heterotic coupling $\lambda^2$ with $A$.

There are generically 24 points on ${\bf P}^1$ where the elliptic
fiber is degenerate.  These 24 points are specified by the
vanishing of the discriminant
$$\Delta=4f^3+ 27 g^2.$$
The points where $\Delta =0$ can also be viewed from the type IIB
viewpoint as the points where we have put 7-branes.  This vacuum
cannot be described in any perturbatively type IIB theory because
the corresponding 7-branes belong to different $(p,q)$ strings
\cite{sc,sch}.

In order to get insight into this duality it is convenient
to have a slightly more detailed relation between the parameters
appearing in the function $f,g$ and the parameters in the heterotic
vacuum.  For the purposes of this paper it suffices to consider
the following restricted class:  Let us consider $E_8\times E_8$
heterotic strings compactified on $T^2$.  Furthermore suppose
we have no Wilson lines turned on and have an unbroken $E_8\times E_8$.
Then the moduli of $T^2$ compactifications of the heterotic theory
are given by the two complex parameters $(\tau,\rho)$ mentioned above.
The F-theory dual is given by the two parameter family $(\alpha ,\beta)$
$$y^2=x^3+\alpha z^4x +(z^5+\beta z^6 +z^7).$$
To show that this is the right family
it suffices to show that we do have $E_8\times E_8$ symmetry.  Assuming
that 7-branes with the ADE singularity give rise to
the corresponding gauge group (a fact
which can be checked perturbatively in type IIB only for the A-series)
we should look for two points on ${\bf P}^1$ where the $K3$ acquires an
$E_8$ singularities.  In the above description the points $z=0$ and $z=\infty$
give rise to $E_8$ singularity.  Let us show this first for the $z=0$
point. In that case the higher powers $\beta z^6+z^7$ in $g$ can be ignored.
Moreover the term $xz^4$ is an irrelevant perturbation
of the singularity $y^2=x^3+z^5$ (as it corresponds to a
trivial element of the chiral algebra) so we have the familiar
$E_8$ singularity (for a discussion in the context of Landau-Ginzburg
theories see \cite{mart,vwa}).  As
for the $z=\infty $ the same argument applies, when we note that
in terms of $\tilde z=1/z$ and the appropriate rescaling
of $x$ and $y$ the story is identical.  In particular the $z^7$
term in the dual variable $\tilde z$ becomes $\tilde z^5$ (recalling
that $g$ has degree $12$) and the first term still looks as $\tilde z^4x$
(as $f$ has degree $8$).  Even though the precise map between $(\alpha,\beta)$
and $(\tau ,\rho)$ can be worked out in terms of the j-functions we
will only be interested in the regime of large $\alpha, \beta$
with $\alpha^3/\beta^2 $ fixed.  In this regime it is easy
to read off that the elliptic fibration becomes essentially a constant
as long as we keep away from 0 and $\infty$.  On the heterotic
side this limit will correspond to large $ \rho$  and moreover the
complex structure $\tau$
of the heterotic $T^2$ gets identified with the complex structure
of the elliptic fiber of the F-theory!
In this limit
we will end up with a geometry of the
7-branes where in addition to the two $E_8$'s on the two sides of ${\bf P}^1$
we have two additional points where the elliptic fiber degenerates---one
near $z=0 $ and the other near
$z=\infty$.  Moreover this implies using the result of \cite{scs} that
the geometry of the ${\bf P}^1$  becomes
elongated looking like
a sausage with the two tips of the sausage corresponding
to $z=0$ and $z=\infty$ respectively.  This follows
from the fact that in this case we have divided the 24 7-branes
into two groups of 12.  Over most
of the region where $z$ is not near $0$ or $\infty$
we simply have a cylinder over which we have a constant elliptic
fiber.  This `cylinder'
smells like a familiar story.  It reminds one of how M-theory gives
rise to $E_8\times E_8$ \cite{hw}.  We will now
argue why this may not be accidental and may be a geometric
realization of string/string duality in the F-theory setup.

One may ask how the type I and heterotic strings can be
realized in the F-theory set up.  Given the M-theory description
of them the idea is as follows (this has also been
noted independently
by \cite{hora}):  Consider M-theory
compactification
on a cylinder down to 9 dimensions.  This gives the heterotic/type I
description in the M-theory setup \cite{hw}. Now if we take the limit
where the volume of the cylinder shrinks to zero we get a type IIB
description of the same object. This is clear when we also recall
that the $\tau$ of type IIB arose from the M-theory side by
compactifying on the two torus to 9 dimensions and taking the small
volume limit of the torus \cite{asp,sc}.
So we would say that
F-theory on a cylinder gives $SO(32)$ type I/heterotic strings
in 10 dimensions.  Note that the fact that F-theory has no local
on shell dynamics in the extra two dimensions make this a consistent
picture \cite{hora}, otherwise locality would have forced us to look
for a product of two groups as the gauge group.  At any rate
the modulus of this cylinder is identified as the coupling
constant of type~I/heterotic strings. Note that this is also in
accord with the fact that after orientifolding type IIB in 10 dimensions
what is left of strong/weak self-duality of type IIB gets mapped
to the strong/weak duality between type I and heterotic strings.

Thus we see that a description of heterotic strings compactified
on $T^2$ down to 8 dimensions can also be viewed as the compactification
of F-theory on a cylinder times $T^2$.  On the other hand the
F-theory/heterotic
duality in 8 dimensions suggests that this is equivalent to
the F-theory compactification on elliptic fibrations over ${\bf P}^1$ which
in the above limit, apart from the tips of ${\bf P}^1$,
becomes  elliptic $T^2$ over the  cylinder.  Moreover the complex
structure of the elliptic fiber $T^2$ is the same as the complex
structure of the $T^2$ of the base on the heterotic side.  The
implication is clear:  string/string duality seems to be the statement of the
exchange symmetry between the base and the fiber in F-theory.
We find this a very appealing picture of how string/string duality
is geometrized in the F-theory set up.

One can also try to develop a similar picture in the M-theory
setup.  Of course by a further compactification on a circle
the same picture repeats, namely, the duality between
M-theory on $K3$ and
heterotic strings on $T^3$ (which themselves can be viewed
as M-theory on $T^3 \times S^1/{\bf Z}_2$) can be seen
as a geometric symmetry of M-theory when we  combine one of the
circles of $T^3$ with $S^1/{\bf Z}_2$ and form a cylinder and
furthermore put the caps at the two ends to make it into a ${\bf P}^1$
as was the case above.  It does however seem that there may also be a
9-dimensional version of the above story, when we think of defining
the analogue of $K3$ in real algebraic geometry.
Consider  the
{\it real}\/ 2-manifold given by the solution to
$$y^2=x^3+f(z)x+g(z)$$
where $x,y,z$ are real and $f,g$ are real polynomials
of degree 8 and 12 respectively.  This can be
viewed as a circle bundle (given by $(y,x)$) over $z$ and in
a suitable limit does look like a cylinder.
It is easy to see that
now we have 18 real parameters to characterize this solution
which could in principle be identified with the
heterotic compactification on $S^1$ (including 16 Wilson
lines, a radius and the coupling constant).
Moreover we can
consider a special limit in which $f=a z^4$ and $g=z^5+bz^6+z^7$.
In this case the two `ends' of the universe are identified
as $z=0,\infty$.  It would be interesting to see if this
picture can be identified directly with M-theory
compactification to 9-dimensions on the above real surface.

\section{$N=1$ F-Theory/Heterotic Duality in 6 Dimensions}

In this section we will rederive some of the results in
\cite{mv} in connection with F-theory duals of heterotic
strings compactified on $K3$.  We will heavily use
 the explicit map discussed in the previous
section.  The methods we will use can clearly be generalized to other
compactifications (say to 4 dimensions).
Our basic strategy is to use the above explicit map to
piece together a certain amount of local data
into a global picture.  The data we need are of two types:
One specifying the fact that we have $K3$ compactification
of the heterotic strings and the other is the data about the bundle.

Let us recall the explicit parametrization we were considering
in the previous section
$$y^2=x^3+\alpha z^4x +(z^5+\beta z^6 +z^7)$$
Recall from the previous section that in the large
$\alpha$ and $\beta$ limit the complex
structure of the torus $y^2=x^3+\alpha z^4x +\beta z^6$ is
the same as that of the complex structure
of the $T^2$ upon which the heterotic string is compactified.
Given this map, the F-theory encoding of the
$K3$ geometry of heterotic string compactification (or other
elliptic compactifications)
becomes straightforward.  We view
the $K3$ on the heterotic side as
compactification of heterotic string on an elliptic fibration
over ${\bf P}^1$ whose affine coordinate we denote by $z'$.
All we have to do is to require
$\alpha $ and $\beta $ now be functions of $z'$ of degree
8 and 12 respectively\footnote{Note that now we have really
20 complex parameters, because the $z=0,
\infty$ are preserved only under the rescaling subgroup
of $SL(2)$ which gives us two additional parameters thus
giving the 20 complex moduli of $K3$ compactification
of heterotic string.}.  We have thus taken care of the
data needed to specify the $K3$ geometry of heterotic
strings in the F-theory setup.

Now we come to the specification of the bundle data.  We are
considering $E_8\times E_8$ bundles.  The simplest method
to try begins by determining what
a single instanton of zero size in an $E_8$
looks like.  Suppose the instanton is localized at $z'=0$.
Then as long as we are away from this point we should have the $E_8$
gauge symmetry restored because after all an instanton of zero
size is simply a singular gauge transformation with singularity
at $z'=0$.  Let us concentrate on the first $E_8$ coming
from the region near $z=0$.
Away from $z'=0$ the elliptic fibration data should be untouched, i.e.,
near $z=0$ but away from $z'=0$ we should get
$$y^2=x^3+z^5$$
The most obvious way to modify this is to consider instead
$$y^2=x^3+z' z^5$$
We conjecture that this is the analogue of a singular gauge transformation
at $z'=0$ in the singularity language.  Note that away from $z'=0$
we have the $E_8$ singularity as desired.  One can immediately
generalize this to having $k_1$ zero size instantons
not necessarily all at the same point, where the locality
of the above description implies we should have
$$y^2=x^3+g_{k_1}(z') z^5$$
where $g_{k_1}$ is a polynomial of degree $k_1$.  For the second
$E_8$ coming from $z=\infty$ the same comments apply and so if
we wish to have instanton number $k_2$ all of zero size at $z'=0$
from the second one
we should have for large $z$ the singularity which goes as
$$y^2=x^3+g_{k_2}(z') z^7.$$

Now let us collect all these facts together and write
the F-theory dual for a heterotic string which has $k_1$
zero size instantons for the first $E_8$ and $k_2$
for the second one on $K3$.  To cancel heterotic
 anomalies (or more precisely if we wish to have no fivebranes
\cite{dmw})
we need to have $k_1+k_2=24$.  It is convenient to write $k_{1,2}=
12{\mp}n$.  With no loss of generality we can take $0\leq n\leq 12$.
  From the above it is thus clear
that the F-theory dual for this configuration is
\begin{equation}\label{geneq}
y^2=x^3+f_8(z')z^4x+g_{12-n}(z')z^5 +g_{12}(z')z^6 +g_{12+n}(z')z^7
\end{equation}
where the subscripts denote the degree of the polynomial.
For this to make sense globally we must take $(z,z')$ not
to parameterize ${\bf P}^1\times {\bf P}^1$ but rather
the rational ruled surface ${\bf F}_n$.
In fact, if we introduce homogeneous partners $w$ and $w'$ for $z$
and $z'$ (and a pair of ${\bf C}^*$'s which act upon them, leaving
the equation invariant),
then in order to ensure that we have a family of $K3$ surfaces,
$f(z,w,z',w')$ and $g(z,w,z',w')$ must be homogeneous of degrees 8 and 12
in $z$ and $w$.  On the other hand, in order to ensure that we don't
get extra instantons at infinity, the polynomials $g_{k_j}(z')$ must
be extended to homogeneous polynomials in $z'$ and $w'$
of the same degree.  Thus,
the homogeneous form of \eqref{geneq} becomes
\begin{equation}\label{geneqhomog}
y^2=x^3+f_8(z',w')z^4w^4x+g_{12-n}(z',w')z^5w^7 +
g_{12}(z',w')z^6w^6 +g_{12+n}(z',w')z^7w^5.
\end{equation}
Taking the ratio of the last two terms in the equation, we find that the
quantity $(z')^nzw^{-1}$ must be invariant under the two ${\bf C}^*$'s.
This means that the ${\bf C}^*$'s acting on $(z,w)$ and $(z',w')$ are
not independent, and that we get ${\bf F}_n$ rather than
${\bf P}^1\times {\bf P}^1$ as the quotient.

Returning to the simpler ``affine'' form of the equation \eqref{geneq},
we can now write the deformation away
from zero size instanton to a finite size (as long as $n$ is not equal
to 9,10,11 where $k_1=3,2,1$ zero size instantons cannot be made to `honest'
finite size instantons on $K3$) in the following form:
\begin{equation}\label{genex}
y^2=x^3+\sum_{k=-4}^{4} f_{8-nk}(z')z^{4-k}x +\sum_{k=-6}^{6}
g_{12-nk}(z')z^{6-k}
\end{equation}
where as usual the subscript denotes the degree of the polynomial
with the condition
that negative degree polynomials are set to zero.

These polynomials give the elliptic Calabi--Yau threefolds with base
being ${\bf F}_n$ which were originally derived in \cite{mv}.  There
is one subtlety to note about this identification: the Weierstrass
models we are using may have singularities.  For example, if we
find a subset of the space of polynomials appearing in \eqref{genex}
for which there are conifold singularities which admit a small resolution,
then the moduli space of type IIA string theory compactified on Calabi--Yau
threefolds will have an entire component corresponding to such polynomials.
In F-theory, though, this should not be considered as a different model:
the additional K\"ahler degrees of freedom provided by the small resolution
are set to zero in F-theory, since they live within the fibers of the
elliptic fibration.  Thus, whereas the moduli space of the 4-dimensional
theories may have several components, their F-theory limits form part
of a space with only one component, completely determined by the base
manifold $B$.

Before considering the various values of $n$ in the above
duality, let us make a remark which will prove helpful in constructing
more dual pairs in lower dimensions (and specifically for $N=1$
vacua in $d=4$).  What we have learned is that
if on the heterotic side we start with a compactification on
an elliptically fibered manifold, the coefficients of $xz^4$ and
$z^6$ specify all the geometry of the manifold. Moreover
if we take the bundle data as $E_8\times E_8$ instantons
of zero size, the sheaf theoretic bundle data can easily be
translated to the geometry of the coefficients $z^5$ and $z^7$.
Once we have constructed this singular limit of the duality
we can extend it in principle in two directions: Either giving
instantons finite sizes, or equivalently viewing the instantons
of zero size as 5-branes, move them to other locations \cite{witfd}.
 The above data is at the intersection
of these two regions and serves as a bridge to either of the two types
of vacua.  The possibility of moving the 5-branes around has been
considered in \cite{sw,han}
and we will consider it further in the F-theory setup in later
sections after we discuss the duality with perturbative heterotic
vacua on $K3$.

\subsection{Determining the gauge group}

Before discussing in detail the various possibilities for $n$, it will
be useful to review how the singularities of the Weierstrass model
(and hence nonabelian part of the gauge group) can be explicitly
determined from a knowledge of the polynomials $f$ and $g$,
where the equation has been written in Weierstrass form
$$y^2=x^3+f(z)x+g(z).$$
If we wish to determine the singularity type at the generic
point of the $z=0$ locus in the base manifold, we should first
calculate the discriminant
$$\Delta(z)=4f(z)^3+27g(z)^2,$$
and then work out the orders of vanishing of $f(z)$,
$g(z)$ and $\Delta(z)$ at $z=0$.  Then according to Tate's algorithm
\cite{tate}, the type of the fiber
on Kodaira's list and the singularity in the Weierstrass model
are given as follows:
\begin{center}
\begin{tabular}{|c|c|c|c|c|} \hline
$\ord(f)$&$\ord(g)$&$\ord(\Delta)$
&fiber-type&singularity-type\\ \hline\hline
$\ge0$&$\ge0$&$0$&smooth&none\\ \hline
$0$&$0$&$n$&$I_n$&$A_{n-1}$\\ \hline
$\ge1 $&$   1  $&$    2 $&$  II $&  none\\ \hline
$  1 $&$   \ge2 $&$   3 $&$  III$&$  A_1$\\ \hline
$ \ge2 $&$  2  $&$    4 $&$  IV $&$  A_2$\\ \hline
$2$&$\ge3$&$n+6$&$I_n^*$&$D_{n+4}$\\ \hline
$\ge2$&$3$&$n+6$&$I_n^*$&$D_{n+4}$\\ \hline
$\ge3$&$  4$  &$  8$&$  IV^* $&$   E_6$\\ \hline
$  3 $&$   \ge5 $&$   9 $&$  III^* $&$  E_7$\\ \hline
$ \ge4$&$   5   $&$   10 $&$  II^* $&$  E_8$\\ \hline
\end{tabular}
\end{center}
If $\ord(\Delta)>10$ then the singularity of the space is so bad
that it destroys the triviality of the canonical bundle on a resolution.

We now turn to a consideration of the various values of $n$ in turn.

\subsection{$n=0,1,2$}

These cases were discussed in detail in \cite{mv}.  In these cases
it is easy to see that for generic polynomials $f$ and $g$,
the elliptic curve \eqref{genex}
is generically non-singular. That the second $E_8$ coming
from $z=\infty$ is generically completely Higgsed is true for all $n$
and this is perfectly consistent with the fact that the above
elliptic curve is non-singular near $z=\infty$
(since all the relevant polynomials
are non-vanishing for the highest powers in $z$,
i.e., the $xz^8$ and $z^{12}$ terms).  For the other $E_8$ coming from
the $z=0$
region it is easy to see that for $n=0,1,2$ the singularity also
disappears in this region.
This is in accord with the fact
that in these cases the gauge groups are completely Higgsed
as can be seen from the heterotic side.
The connection between
$n=0$ and $n=2$ has been noted in detail in \cite{mv} where it is noted
that they are in fact the same theory (similar remarks appear in \cite{ibet}).

\subsection{$n=3$}

In all the cases $n>2$ the first $E_8$ cannot be completely Higgsed
and we are left with some gauge group.
In this case the most relevant terms near $z=0$ are $f_2$ and $g_0$
but since  $xz^2$ is subleading to $z^2$ in the small $(x,z)$ region
the singularity is that of
$$y^2=x^3+ z^2$$
which is the $SU(3)$ singularity, so in this case we have a generic
unbroken $SU(3)$ without matter (i.e., the singularity is of an ordinary
type and the 7-brane analysis suffices to conclude we have no matter).
This is in accord with the observation \cite{dmw} that in this case
there is a branch with a generic unbroken $SU(3)$.

\subsection{$n=4$}

This is a very interesting case, as it has been conjectured
in \cite{mv} that not only does this correspond to $E_8\times E_8$ with
instanton numbers $(8,16)$ but also to heterotic (or type I)
$SO(32)$ string on $K3$ (see also \cite{ewa}).
One check for this is that both the $(8,16)$
$E_8\times E_8$
and $SO(32)$ heterotic strings leave an unbroken $SO(8)$ generically.
This also follows from considering $SO(8)\times SO(16)$
bundles on $K3$ of instanton numbers $(8,16)$.  This is a subbundle
of both $SO(32)$ {\it and}\/ $E_8\times E_8$ bundles of total
instanton number 24 which resides as (8,16) in the $E_8\times E_8$
subgroup.  Clearly this leaves an $SO(8)$ subgroup of
the first $E_8$ unbroken.  Moreover we expect no matter in this case.
Now let us check whether
the generic singularity of the F-theory dual is of this type.
In this case the two lowest terms in \eqref{genex} correspond
to $f_0 xz^2+g_0z^3$.  Both of these terms are of the
same order and the generic singularity can be written
as
$$y^2=x^3+z^3 \sim x^3+xz^2$$
and is recognized as that of $D_4$, i.e., $SO(8)$.

\subsection{$n=5$}

{}From the heterotic side we expect an $E_6$ generic singularity
with a charged hypermultiplet in the $27$ (in addition to
certain number of neutral hypermultiplets).  In the F-theory
description the leading term in \eqref{genex} is $f_3(z')xz^3+g_2(z')z^4$
and thus the second term dominates and we have for the leading singularity
$$y^2=x^3+g_2(z')z^4.$$
Note that generically this is the singularity for $E_6$ and so
we should expect an $E_6$ gauge group.  However the fact that we
have at the zeros of $g_2(z')$ a different singularity should be
interpreted, in light of the above result in the heterotic
string, as a signal of extra matter and in fact of a single 27 hypermultiplet.

\subsection{$n=6$}

This is similar to the case $n=5$ except that now on the heterotic
side we expect to get $E_6$ with no extra matter and this is reflected
by the fact that the leading singularity is now given by
$$y^2=x^3+z^4$$
which is the $E_6$ singularity without any extra singularities
at special values of $z'$.

\subsection{$n=7$}

In this case the leading terms in \eqref{genex} come from
$f_1(z') xz^3+z^5 g_5(z')$ and the $xz^3$ term is more relevant
than the $z^5$ term and so we get
for the leading singularity
$$y^2=x^3+f_1(z') xz^3$$
which is generically the $E_7$ singularity. Again since
there is an extra singularity in the above when $f_1$
vanishes we expect extra charged matter.  This is in accord
with the heterotic string analysis where one ends up with $E_7$
gauge symmetry with a $1/2$ hypermultiplet of 56.  Thus we learn
through this example to recognize the above singularity
as corresponding to an $E_7$ with half a 56 hypermultiplet.
It would be interesting to verify this
also from the analysis of monodromies and vanishing cycles
as in \cite{bsv}.

\subsection{$n=8$}

This is very similar to the case above except that the relevant
term is $f_0 xz^3$ and so we end up with
$$y^2=x^3+xz^3$$
which is an ordinary $E_7$ singularity and we should have no
extra charged matter.  This is in accord with the heterotic
string result.

\subsection{$n=9,10,11$}

In these cases the leading singularity is that of $E_8$
as the $z^5$ term is the more relevant term (except that
in all cases we have extra singularities).  On the heterotic
side these correspond to zero size instantons which cannot
be made to finite size ones, so the physics is somewhat
singular.   In fact these correspond to transition points
noted above.  Unlike the above cases
where the extra singularity signaled the existence
of extra matter, this does not seem to have a well defined meaning in
this case as one sees
from the heterotic side.
These transitions will be discussed further below.

\subsection{$n=12$}

This is the case where the leading singularity is pure
$E_8$ with no extra singularity.  So we expect ordinary $E_8$.
This is in accord with the fact that in this case all the instanton
numbers are on the other $E_8$ bundle in the heterotic picture.

\subsection{Matter from extra singularities}

We have only explored a very tiny piece of information
encoded in \eqref{genex}.  For example, we can learn about how
various extra singularities correspond to extra matter.
We have seen some examples above and we can and should be
more systematic.
Just to give a flavor of what one means consider the cases
above where we have the generic gauge group Higgsed to
a subgroup of $E_7$.  In this case by tuning appropriate
moduli we can end up at special points where we have
an extra $E_7$ singularity.  For example tune the parameters
so that the leading term comes from $f_{8-n}xz^3$, and thus
the leading term is given by
$$y^2=x^3+f_{8-n}(z')xz^3$$
{}From the heterotic side this corresponds to instanton number
$12-n$ in the $E_7$ and we should expect to find ${1\over 2}k_1 -2=
{1\over 2}(8-n)$
charged hypermultiplets in the 56 of $E_7$.  This supports the
above result that each time we see a zero in front of the $xz^3$
term this is a source of a half 56 of $E_7$ as was also found above.
Had we tuned to obtain $E_6$ singularities instead we would
have found
$$y^2=x^3+g_{12-2n}(z')z^4.$$
{}From the heterotic side by further Higgsing the above $E_7$ to
$E_6$ where we lose a 56 in the process of Higgsing and
each surviving 56 turns into a pair of 27's we learn that
we would expect $k_1-6=6-n$ charged 27's.
This is again consistent with the fact that the
coefficient of the above singularity is a polynomial
of degree $2(6-n)$ in $z'$ and suggests that each pair of zeroes in
front of $z^4$ term
gives rise to a $27$.
Clearly by following various different branches of Higgsing
we end up with various groups with varying amount of matter
and we can thus begin to build up the dictionary of what
matter representation corresponds to what kinds of singularity.

The geometric interpretation of this matter content should be this:
along each curve $\Sigma_i$ in the base where the elliptic curves
become singular, there is a generic singularity type but there are
also finitely many points $P_{ij}$ where the singularity type changes.
These may arise through intersections with other components $\Sigma_k$,
or simply because of an extra vanishing of the polynomials $f$ and/or
$g$ at those points.  There is a vast array of possibilities for
what kinds of singularities can arise; these have been classified
mathematically
\cite{miranda,grassimin} under the assumption that the $\Sigma_i$'s
meet transversally, but that might not always be the case.  The simplest
example of this phenomenon is the case of a curve of $A_k$ singularities
meeting a curve of $A_\ell$ singularities, where a D-brane interpretation
for the matter was given in \cite{bsv}.

\section{General Remarks about F-Theory on Calabi--Yau 3-Folds}

In this section we discuss some generalities about what
spectrum to expect for F-theory compactifications on Calabi--Yau 3-folds.
Consider an elliptic Calabi--Yau 3-fold $X$ with
Hodge numbers $h^{1,1}(X)$
and $h^{2,1}(X)$.   Note furthermore that we also have a base $B$
which has a certain Hodge number $h^{1,1}(B)$. Note that
$h^{1,1}(B)+1$ and $h^{1,1}(X)$ need not be equal
as there may be some K\"ahler deformations of the Calabi--Yau manifold which
do not correspond to the Calabi--Yau being elliptically fibered.

We would like to determine how many tensor multiplets $ T$,
vector multiplets $V$ and hypermultiplets $H$ we have.
The simplest thing to work out is the number of tensor multiplets
$T$. The scalars in these
multiplets are in one to one correspondence with the K\"ahler
classes of $B$ except for the overall volume of $B$ which corresponds
to a hypermultiplet \cite{vft}.  We thus have
\begin{equation}\label{nten}
T=h^{1,1}(B)-1
\end{equation}
In order to obtain further information about the rest of the
degeneracies we use the fact that
upon further compactification on $T^2$ our compactified
F-theory model becomes equivalent to the
type IIA theory compactified
on the same Calabi--Yau \cite{vft}.  We thus learn,
by going to the Coulomb phase of the vector multiplets in the 4d sense, that
\begin{equation}\label{cons}
r(V)+T=h^{1,1}(X)-2
\end{equation}
\begin{equation}\label{conh}
H^0=h^{2,1}(X)+1
\end{equation}
where $r(V)$ denotes the rank of the vector multiplet and $H^0$
denotes the number of neutral hypermultiplets which are uncharged
with respect to the Cartan of $V$.  Combining the above with \eqref{nten}
we have
\begin{equation}\label{bes}
T=h^{1,1}(B)-1; \qquad r(V)=h^{1,1}(X)-h^{1,1}(B)-1; \qquad
H^0=h^{2,1}(X)+1
\end{equation}
We can determine the non-abelian factors in $V$ simply
by studying the singularity type of the 7-branes
(and if we had a better understanding of how various
matter representations correspond to various type of singularities
we could also have determined H in this way).  We can
thus deduce the number of $U(1)$'s because from \eqref{bes}
we know the total rank $r(V)$ of the gauge group.

On the other hand, for each non-abelian factor in $V$ the
rank of that factor is one less than the number of components of the fiber
lying over the corresponding curve in the base manifold.  Geometrically,
the elements of $H^{1,1}(X)$ must come from divisors in the base
(i.e., $H^{1,1}(B)$), ``extra''
components of the reducible fibers (which contribute a total of
$r(V')$ to $h^{1,1}(X)$, where $V'$ is the nonabelian part of the
gauge group), and sections of the fibration.  Comparing
this with the spectrum of the F-theory, we see that
the rank of the group of sections\footnote{To determine the rank of the
group of sections, pick a generic fiber $E$ of the elliptic fibration,
and consider the intersection of all sections of the fibration with
$E$. (Each section determines a single point of $E$.)  This collection of
points is a subgroup of $E$ which is known to be finitely generated;
the {\it rank}\/ refers to its rank as an abelian group, i.e., the
group is ${\Z}^{\mbox{rank}}\times(\mbox{finite order})$.  Note that if there
is exactly one section, it must be the `zero' of the group structure
on $E$, and the rank is $0$.}
 precisely
gives the number of $U(1)$ factors in the gauge group.
Roughly speaking
what this means is that the number of $U(1)$'s is the number
of ways we can choose the base manifold $B$ inside the
elliptic Calabi--Yau.

The above constraints (together
with the fact that the representations
of non-abelian groups are typically rather constrained)
are usually powerful enough
to also fix $H$.  There is a
constraint  from anomaly cancellation condition
which requires that \cite{anoc}
\begin{equation}\label{anc}
H-V=273-29 T
\end{equation}
where $H$ and $V$ denote the total number of hypermultiplets
and vector multiplets respectively. This turns out to be
a strong check on the spectrum of matter in the F-theory
compactification discussed above.

\section{The Structure of the Base}

It is known \cite{grassi,grassibis,oguiso}
that for any elliptic fibration $\pi:X\to B$ of a
Calabi--Yau threefold $X$, the base surface $B$ has at worst orbifold
singularities.  In fact, the singularities are more constrained than
that: together with the collection of curves $\Sigma_i$ which
specify where the elliptic fibration is singular, the singularities
have a special property known as ``log-terminal''.  There are in
fact not many surfaces which can occur as the base \cite{grassibis}.

For our present purposes, we shall use the following properties of $B$.
First of all, if the singularities of $B$ are resolved then $B$ is
either a surface with $12K_B=0$ (i.e., a $K3$ surface, Enriques surface,
hyperelliptic surface, or complex torus), or $B=(T^2\times {\bf P}^1)/G$,
or (the most common case) $B$ is a blowup of ${\bf P}^2$ or ${\bf F}_n$.
In the case that $B$ is a $K3$ surface, a complex torus, a hyperelliptic
surface, or $(T^2\times {\bf P}^1)/G$, the corresponding Calabi--Yau
will have a holomorphic one-form or a holomorphic two-form, and hence
the corresponding type II compactification has extra supersymmetry.  We shall
exclude these cases from consideration.

Thus, for our purposes, we can take $B$ to be either an Enriques surface,
a blowup of ${\bf P}^2$ or ${\bf F}_n$, or a surface with orbifold
singularities whose resolution of singularities is one of the above.

\section{Examples}

In this section we discuss various examples of
F-theory compactifications on Calabi--Yau threefolds.
The first class of examples we consider correspond
to the case where the base manifold is smooth, and should
thus be, according to our previous discussion ${\bf P}^2$
or ${\bf F}_n$, blown up at various number of points.
In the second class we find a large range of possibilities for the
Hodge numbers.
In the third class we explore the possibility of having
orbifold singularities in the base.  The fact that what
we find seems self-consistent suggests that at least orbifold
singularities of the base may be harmless for F-theory compactifications.

\subsection{Blown up ${\bf F}_n$ as the base}

Each of the ${\bf F}_n$ models reviewed in section 3 can be blown
up to yield further models.  We illustrate this by performing the blowup at
$z=z'=0$.  The simplest way to do this is to introduce a new coordinate
$u$, along with a ${\bf C}^*$ which acts on the coordinates as
$$(x,y,z,z',u)\mapsto
(\lambda^2x,\lambda^3y,\lambda z,\lambda z',\lambda^{-1}u)$$
for $\lambda\in{\bf C}^*$, and to remove the locus where $z=z'=0$
before forming the quotient by this action.  When done in this way,
the two coordinate charts which are usually used to describe a blowup
are quite visible: when $z\ne0$ we have
$$(x,y,z,z',u)\sim(z^{-2}x,z^{-3}y,1,z^{-1}z',zu)$$
and we should use those four quantities as coordinates; the other chart
with $z'\ne0$ is obtained similarly.

However, in addition to a complex structure we should specify a value
for the new K\"ahler class on the blowup. This is done by choosing
a positive number $r$, and then describing the quotient by first imposing
the constraint
$$\frac12(2|x|^2+3|y|^2+|z|^2+|z'|^2-|u|^2)=r,$$
and then taking the quotient by $U(1)$ only.  Notice that in this
formulation,
the removal of the locus $\{z=z'=0\}$ is automatic.

To determine the equation for the Calabi--Yau threefolds in this space, we
simply
supplement each term in \eqref{genex} with an appropriate power of
$u$ in order to guarantee invariance under the new ${\bf C}^*$,
obtaining:
\begin{equation}\label{genexbl}
y^2=x^3+\sum_{k=-4}^{4}\sum_{\ell=0}^{8-nk} f_{k\ell}\,z'{}^\ell z^{4-k}
u^{\ell-k}x
+\sum_{k=-6}^{6}\sum_{\ell=0}^{12-nk} g_{k\ell}\,
z'{}^\ell z^{6-k}u^{\ell-k}.
\end{equation}
Certain of the terms from the original equation---those with
$\ell<k$ for $k\ge0$---must
be suppressed in the blown up model in order to obtain
a Calabi--Yau hypersurface there.  (That is, we must have
$z'{}^k|f_{8-nk}(z')$ and $z'{}^k|g_{12-nk}(z')$ for $k\ge0$.)
So as expected, we find that $h^{1,1}$
increases and $h^{2,1}$ decreases during this type of blowup.

To generalize this construction somewhat, we can blow up a collection
of points $z=0$, $z'=a_i$, for $i=1,\dots,r$.  If we let
$\varphi(z')=\prod (z'-a_i)$, then the condition on the equation
becomes: $\varphi(z')^k|f_{8-nk}$, $\varphi(z')^k|g_{12-nk}$.
This gets increasingly difficult to satisfy as $r$ increases.
However, we can go as high as $r=24$ if we take $n=-12$ and choose
our equation of the form
$$y^2=x^3+f_8(z')z^4+\varphi(z')z^5+g_{12}(z')z^6+z^7,$$
with unbroken $E_8\times E_8$.  More generally, starting from any value
of $n$ we can blow up as many as $12+n$ points along $z=0$ and $12-n$ points
along $z=\infty$ by using a similar construction.

If we try to blow up points at arbitrary locations, however, it becomes
increasingly difficult to verify that the terms suppressed in the
newly blown up model do not cause unintended additional singularities
in the space---singularities which could cause the Calabi--Yau condition to
be violated.
The blowing up of 24 points, which we were able to do when the points
were in very special position, cannot be done if the points are generic.

So we will adopt a different strategy for producing examples: first we
build models in which the base is already blown up a large amount,
and then we study blow downs (by means of extremal transitions).

\subsection{Blown up ${\bf P}^2$ as the base}

If we blow up ${\bf P}^2$ at the 9 points of intersection of two
cubic curves, we obtain a rational surface with an elliptic fibration
(and a section).  Generically there will be 12 singular fibers in such
a fibration.  Let $Y_1$ and $Y_2$ be two such blown-up ${\bf P}^2$'s.
We can use the elliptic fibrations $\pi_i:Y_i\to{\bf P}^1$ to
define a Calabi--Yau threefold, as follows:
$$X=\{(y_1,y_2)\in Y_1\times Y_2\ |\ \pi_1(y_1)=\pi_2(y_2)\}.$$
(This is called the ``fiber product'' construction, and was studied
extensively in \cite{schoen}.)  The threefold
$X$ maps to $Y_2$ as a base manifold, and the fibers of that map
are $T^2$'s whose complex structure is determined by the fibration $Y_1$.

The Hodge numbers are $h^{1,1}(X)=h^{2,1}(X)=19$, and all of the
moduli can be obtained by varying the fibrations $Y_1$ and $Y_2$.
This model was briefly discussed in \cite{vft}, and has
9 tensor multiplets 8 vector multiplets and 20 hypermultiplets.

As shown in \cite{schoen}, and
as we will see later, many more examples (with smaller values of
$h^{1,1}$) can be produced from this one by extremal transitions.

\subsection{The Voisin--Borcea examples}

There is a rich class of elliptic Calabi--Yau threefolds
studied independently by Voisin \cite{voisin} and Borcea \cite{borcea}
and recently discussed by Aspinwall \cite{AspinExtreme}.
Start with a $K3$ surface $S$ which admits an involution $\sigma$
such that $\sigma^*(\omega)=-\omega$, where $\omega$ is the
holomorphic $2$-form.  Build a Calabi--Yau manifold of the form
$S\times T^2/(\sigma,-1)$, resolving singularities appropriately.
These are the Voisin--Borcea models.  They are elliptically fibered
over a base $B=S/\sigma$, and so can be used for F-theory compactification.

\begin{figure}
\setlength{\unitlength}{0.008in}%
$$\begin{picture}(445,266)(60,385)
\thinlines
\put(100,420){\circle*{6}}
\put(140,420){\circle*{6}}
\put(260,420){\circle*{6}}
\put(300,420){\circle*{6}}
\put(420,420){\circle*{6}}
\put(460,420){\circle*{6}}
\put(120,440){\circle*{6}}
\put(160,440){\circle*{6}}
\put(240,440){\circle*{6}}
\put(280,440){\circle*{6}}
\put(320,440){\circle*{6}}
\put(400,440){\circle*{6}}
\put(440,440){\circle*{6}}
\put(140,460){\circle*{6}}
\put(180,460){\circle*{6}}
\put(220,460){\circle*{6}}
\put(260,460){\circle*{6}}
\put(300,460){\circle*{6}}
\put(340,460){\circle*{6}}
\put(380,460){\circle*{6}}
\put(420,460){\circle*{6}}
\put(160,480){\circle*{6}}
\put(200,480){\circle*{6}}
\put(240,480){\circle*{6}}
\put(280,480){\circle*{6}}
\put(360,480){\circle*{6}}
\put(400,480){\circle*{6}}
\put(180,500){\circle*{6}}
\put(220,500){\circle*{6}}
\put(260,500){\circle*{6}}
\put(300,500){\circle*{6}}
\put(340,500){\circle*{6}}
\put(380,500){\circle*{6}}
\put(200,520){\circle*{6}}
\put(240,520){\circle*{6}}
\put(280,520){\circle*{6}}
\put(320,520){\circle*{6}}
\put(360,520){\circle*{6}}
\put(220,540){\circle*{6}}
\put(260,540){\circle*{6}}
\put(300,540){\circle*{6}}
\put(340,540){\circle*{6}}
\put(240,560){\circle*{6}}
\put(280,560){\circle*{6}}
\put(320,560){\circle*{6}}
\put(260,580){\circle*{6}}
\put(300,580){\circle*{6}}
\put(280,600){\circle*{6}}
\put(320,480){\circle*{6}}
\put(120,400){\circle{10}}
\put(280,400){\circle{10}}
\put(440,400){\circle{10}}
\put(200,440){\circle{10}}
\put(200,480){\circle{10}}
\put(360,440){\circle{10}}
\put(360,480){\circle{10}}
\put(280,480){\circle{10}}
\put(280,520){\circle{10}}
\put(280,560){\circle{10}}
\put(280,600){\circle{10}}
\put(280,440){\circle{10}}
\put(120,440){\circle{10}}
\put(440,440){\circle{10}}
\put( 80,400){\line( 1, 0){420}}
\put( 80,400){\line( 0, 1){240}}
\put(300,620){\circle*{6}}
\put(320,600){\circle*{6}}
\put(340,580){\circle*{6}}
\put(380,540){\circle*{6}}
\put(400,520){\circle*{6}}
\put(420,500){\circle*{6}}
\put(440,480){\circle*{6}}
\put(440,480){\circle{10}}
\put(460,460){\circle*{6}}
\put(480,440){\circle*{6}}
\put(360,520){\circle{10}}
\put(360,560){\circle*{6}}
\put( 75,379){\makebox(0,0)[lb]{\raisebox{0pt}[0pt][0pt]{0}}}
\put(175,379){\makebox(0,0)[lb]{\raisebox{0pt}[0pt][0pt]{5}}}
\put(270,379){\makebox(0,0)[lb]{\raisebox{0pt}[0pt][0pt]{10}}}
\put(370,379){\makebox(0,0)[lb]{\raisebox{0pt}[0pt][0pt]{15}}}
\put(470,379){\makebox(0,0)[lb]{\raisebox{0pt}[0pt][0pt]{20}}}
\put( 65,495){\makebox(0,0)[lb]{\raisebox{0pt}[0pt][0pt]{5}}}
\put( 55,595){\makebox(0,0)[lb]{\raisebox{0pt}[0pt][0pt]{10}}}
\put(505,395){\makebox(0,0)[lb]{\raisebox{0pt}[0pt][0pt]{$r$}}}
\put( 75,645){\makebox(0,0)[lb]{\raisebox{0pt}[0pt][0pt]{$a$}}}
\end{picture}$$
	\caption{The Voisin--Borcea examples.}
	\label{figVB}
\end{figure}
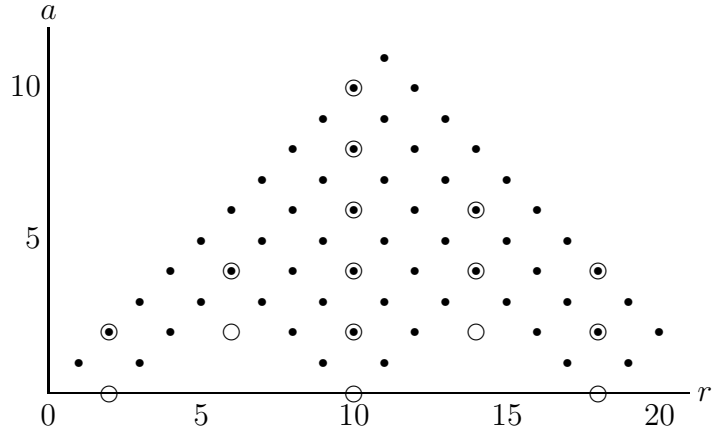

One way to produce such $S$'s and $\sigma$'s is to take one of the
two ${\bf Z}_2$'s in our ${\bf Z}_2\times {\bf Z}_2$
orbifolding constructions, and
mod out $T^4$ by it to produce a Kummer surface.  But there are
many more such pairs, all classified by Nikulin \cite{nikulin}
in terms of three invariants $(r,a,\delta)$.  These examples are
plotted in figure \ref{figVB} according to the values of $r$ and $a$; the open
circles
represent cases with $\delta=0$ (which means that the fixed locus of
$\sigma$
defines a class in $H^{1,1}(S,{\bf Z})$ which is divisible by $2$)
and the closed circles represent
cases with $\delta=1$ (which means that the class of the fixed locus
is not divisible by 2).  In the case of invariants $(10,10,0)$, the
surface $S/\sigma$ is an Enriques surface and the Hodge numbers of the
corresponding threefold are $h^{1,1}=h^{2,1}=11$.  In all other
cases, $S/\sigma$ is a rational surface, and
the Hodge numbers of the corresponding
Calabi--Yau threefold are
$$h^{1,1}=5+3r-2a$$
$$h^{2,1}=65-3r-2a.$$

When $(r,a,\delta)=(10,10,0)$, $\sigma$ has no fixed points and
$S/\sigma$ is an Enriques surface.  When $(r,a,\delta)=(10,8,0)$,
the fixed points of $\sigma$ consist of two curves of genus $1$.
(In fact, in this case the Hodge numbers are (19,19) and we have
the blown up ${\bf P}^2$ model discussed in the previous subsection.)
In all other cases,
the fixed points of $\sigma$ consist
of a curve $C$ of genus $g$ together with $k$ rational curves
$E_1$, \dots, $E_k$, where
$$g={1\over2}(22-r-a)$$
$$k={1\over 2}(r-a).$$
The elliptic fibration over $B$ must have fibers of Kodaira type $I_0^*$
along each component of the fixed locus, and the Weierstrass model
will have $D_4$ singularities over those components.  To determine
the structure of $B$, recall that the canonical bundle formula for
an elliptic fibration then guarantees that
$$-K_B=\frac12 C+\sum_{i=1}^k\frac12 E_i,$$
from which it follows that $c_1^2(B)=\frac14(4g-4 + (-4)k)=10-r$.
Since $c_1^2(B)+c_2(B)=12$, we find that $c_2(B)=2+r$ and so
$h^{1,1}(B)=r$.  The number of tensor multiplets ranges from a
low of 0 in the case $(r,a,\delta)=(1,1,1)$ (in which case $B={\bf P}^2$),
to a high of 19 in the case $(r,a,\delta)=(20,2,1)$.  Note that there
are also several familiar\footnote{These models differ slightly from
the familiar ones, due to extra singularities of the Weierstrass form,
as discussed in section 3.} models with exactly one tensor multiplet:
$(2,0,0)$ which corresponds to ${\bf F}_4$, $(2,2,1)$ which corresponds
to ${\bf F}_1$ and $(2,2,0)$ which corresponds to ${\bf F}_0$.

Let us verify the anomaly formula for these models (in the cases
$(r,a,\delta)\ne(10,10,0), (10,8,0)$).  There is an enhanced
gauge symmetry of $SO(8)^{k+1}$, one factor of $SO(8)$ coming from
each component of $-K_B$.
On the other hand, from \eqref{bes} we find
$$r(V)=(5+3r-2a)-r-1=4(k+1),$$
so there are no additional $U(1)$ factors in the gauge group.  It follows
that $V=28(k+1)$.

The factor of $SO(8)$ associated to $C$ has a matter representation
consisting
of $g$ adjoints, and so contributes $(28-4)g$ charged hypermultiplets
to the spectrum.
There is no matter coming from the other factors, since they are associated
to rational curves.  We therefore find that
$$H=(65-3r-2a)+1+24g,$$
and can verify the anomaly formula \eqref{anc}:
$$H-V=(65-3r-2a)+1+24(\frac{22-r-a}2)-28(\frac{r-a}2+1)
=273-29(r-1).$$

\subsection{Orbifold examples}

Some of the Voisin--Borcea examples have a simple orbifold description,
which we will now discuss in more detail.  Moreover we discuss
also the Z-orbifold which is not among the Voisin--Borcea
manifolds but presents a potentially interesting F-theory compactification.

If we consider an M-theory compactification on an elliptic
orbifold and take the limit in which the K\"ahler class of the
elliptic fiber goes to zero, we would expect to obtain in the
limit the F-theory on the same orbifold.  Thus we immediately
are led to accept the existence of orbifold singularities in the
base as harmless.  In this section we consider 4 orbifold examples.
Three of them correspond to ${\bf Z}_2\times {\bf Z}_2$ orbifolds and
one corresponds to a ${\bf Z}_3$ orbifold giving rise to the Z-manifold.

Consider the Calabi--Yau threefold given by the orbifold of $T^2\times
T^2\times T^2$.  Let the coordinates of the tori be denoted
by $z_1,z_2,z_3$.  In the ${\bf Z}_2\times {\bf Z}_2$ examples we have
the two generators flipping the sign of two out of three coordinates.
Depending on whether they are accompanied by any extra shifts
in the coordinates we end up with three possibilities.
Let $g_{ij}$ denote the flipping of sign of $z_i$ and $z_j$.
Then the three cases are:

(i) The two generators are
$g_{12}$ accompanied by a shift of order 2 in $z_3$
and $g_{13}$ accompanied by a shift of order 2 in $z_2$.
This manifold has Hodge numbers (11,11) (and has appeared
in the context of type II/heterotic duality in \cite{secq}).

(ii) Same as above except $g_{13}$ is not accompanied by a shift.
This manifold has Hodge numbers (19,19).

(iii) Same as above except with no shifts at all.  This manifold
has Hodge numbers $(h^{1,1},h^{12})=(51,3)$.

Except for the last case the other cases can be chosen
to be F-theory vacua in two different ways depending on which
$T^2$ we choose as the elliptic fiber.  Here we consider the
simplest possibility where the first $T^2$ is the elliptic
fiber (the other cases have some interesting features
which we have not fully understood).

In case (i) the base $B$ is the Enriques surface and
has $h^{1,1}(B)=10$ so according to \eqref{bes} we have $T=9$,
$r(V)=0$, $H=12$.  Note that the anomaly formula \eqref{anc} is
automatically satisfied.
In this case the manifold can be deformed
from the singular orbifold limit to a smooth manifold which
still respects the elliptic fibration.
This model is a Voisin--Borcea model with $(r,a,\delta)=(10,10,0)$.

In case (ii) the base manifold turns out to be the same
as that studied in the previous section:  ${\bf P}^2$ blown
up at nine points.   We thus have generically\footnote{At
the orbifold point we end up with $SO(8)^2$ with an adjoint
matter which can be deformed to the above spectrum by going
away from the orbifold limit.}
$U(1)^8$ and
$T=9$, $H=12$.  Again this manifold can be deformed
away from the singular limit where the base is smooth.
This model is dual to the models studied in the M-theory
setup in \cite{sen} and as a type IIB orientifold in \cite{dab}.
This is the same model as the Voisin--Borcea model with
$(r,a,\delta)=(10,8,0)$.

Case (iii) is more interesting in that the gauge group is nonabelian
even at generic values of the moduli.  This model coincides
with the Voisin--Borcea model with $(r,a,\delta)=(18,4,0)$,
and we have verified the anomaly formula for it in the previous
subsection.

Note that the divisor on the base manifold can be accounted for as
$h^{1,1}(B)=16+1+1$, where 16 come from the fixed point
of the group element $g_{23}$ and $1+1$ come from the K\"ahler
classes of the second and third $T^2$.  We thus find from \eqref{bes}
that $T=17$.

Finally we consider the case of the Z-orbifold which
is given by modding $T^2\times T^2\times T^2$ with
a ${\bf Z}_3$ which acts on each $T^2$ by $z\rightarrow
{\rm exp}(2\pi i/3)z$ (and we are assuming that each
$T^2$ does have the ${\bf Z}_3$ symmetry).
The K\"ahler moduli of the base, compatible with elliptic fibration
is 4 dimensional and we thus learn that $T=3$.  Moreover
since $(h^{1,1},h^{2,1})=(36,0)$ we learn from \eqref{bes} that
$r(V)=31$ and $H^0=1$.  However the description of the low
energy degrees of freedom are not just these. We will have
tensionless strings, as we will discuss further in the next section.
Note that
this is the case where
the physics is stuck at a place with tensionless strings where
we cannot make a transition to other $N=1$ vacua and the vacuum
appears to be isolated. This seems to indicate that $N=1$ vacua
do not form a connected set in 6 dimensions.  Note that the
fact that the anomaly formula is not satisfied is another
indication that we are missing certain states, and the local
Lagrangian description does not suffice to describe this vacuum.

\section{Transitions Among $N=1$ Vacua}

We have constructed various examples of $N=1$ vacua in $d=6$
in the F-theory setup.  It is natural to wonder whether
there are transitions among them.  Let us recall that in the
F-theory setup the compactification is on an elliptically fibered
Calabi--Yau threefold and that upon further compactification
to 5 and 4 it is on the same moduli as M-theory and type IIA
on the same manifold.  If there is a transition among the vacua
in 6-dimensions, there will clearly continue to be transitions
upon further compactifications.  This will give us a handle on
various types of transitions as there are a number of tools
available to study transitions
among Calabi--Yau threefolds.
However one has to be somewhat careful as the converse
may not be true; in other words we may have transitions
in 4d or 5d vacua which do not occur in 6d.  One can
also in principle imagine transitions among 4d vacua
which do not give transitions in 5d, if one uses a finite
value of the coupling constant of type IIA (which is related
to the radius of the extra circle from 5 to 4) or the complexified
K\"ahler class to make the
transition.

Now suppose we start with type IIA compactification to four
dimensions on a Calabi--Yau threefold which is elliptically
fibered.  If we consider a transition to another
elliptically fibered Calabi--Yau and note that such a transition
is insensitive to the size of the elliptic fiber
or the coupling constant of type IIA, then in the limit
of large coupling constant and zero size for the elliptic
fiber we recover the F-theory transition in 6 dimensions.
There are, however, transitions which connect elliptically
fibered Calabi--Yau manifolds to non-elliptically fibered ones.
These will not necessarily lead to transition in 6d.  Thus even though
the space of Calabi--Yau threefolds seems to be connected through
extremal transitions, the same cannot be said about elliptically
fibered ones, and in particular about the connectivity of $N=1$
vacua in 6d.  An example of this is the Z-orbifold considered before
which is not connected to any other Calabi--Yau unless we go through regions
where the manifold is not elliptically fibered, i.e., we have
to blow up the orbifold singularities which destroys elliptic
fibration as we try to go to transition points to other Calabi--Yau manifolds.
In the following we will divide our discussion into two parts, one involving
the mathematical aspects of transitions, which is much
better understood, and the other discussing the physical interpretation
of such transitions.

\subsection{Mathematical aspects of threefold transitions}

The type IIA theory compactified on a Calabi--Yau threefold can undergo
an extremal transition leading to another branch of the moduli space.
We discuss these transitions from a mathematical perspective in
this subsection.

In the semiclassical approximation, if we vary the K\"ahler moduli
of a Calabi--Yau threefold $X$ so as to reach a boundary wall of the
K\"ahler cone of $X$, some portion of $X$ shrinks to zero size.
The precise ways in which this can happen are known from a mathematical
analysis of the K\"ahler cone \cite{wilson}: at a codimension one point of
the boundary, we can either have\footnote{There is one further possibility
which has not yet been excluded in the mathematics literature---a curved
portion of the boundary of the K\"ahler cone---but no examples of this
phenomenon are known.}

(i) the fibers of a $K3$ fibration or a $T^4$ fibration 
shrink to zero size, or

(ii) the fibers of an elliptic fibration shrink to zero size, or

(iii) a 2-cycle shrinks to a point, or

(iv) a 2-parameter family of 2-cycles shrink to points (i.e., a 4-cycle
shrinks down to a 2-cycle), or

(v) a 4-cycle shrinks to a point.

\noindent
Note that worldsheet corrections will affect the location of the
singularity in the moduli space, but not its qualitative nature.

According to Hayakawa's criterion \cite{hayakawa},
cases (i) and (ii) occur at points at infinite distance in the moduli
space,
while cases (iii), (iv) and (v) occur at finite distance.
Case (iii) includes conifold points, where Strominger \cite{andy} showed
how
massless solitons cure the singularity in the physics
(albeit from the mirror, IIB perspective).  Case (iv) corresponds to
enhanced nonabelian gauge symmetry \cite{AspinExtreme,bsv}
where a similar analysis
\cite{KMP} shows
how the physics gets corrected nonperturbatively.  Case (v) is
relatively unexplored, although examples have been discussed in \cite{CGGK}.

In the conifold case \cite{GMS} and in the enhanced gauge symmetry case
\cite{KMP},
new branches of the moduli space can arise if the singularities have
appropriate properties.  Mathematically, at least,
the same is true in case (v) in the examples discussed in \cite{CGGK}.

The singularities which are introduced in case (v) when a 4-cycle shrinks
to a point are of a special type known in the mathematics literature
as ``isolated canonical singularities with a crepant blowup''.
A rough classification of
these was given many years ago \cite{reid}, based on the structure of
the 4-cycles which occur.  Such a 4-cycle must be a special type of
complex algebraic surface known as a {\it generalized del Pezzo surface};
the adjective ``generalized'' refers to the fact that the surface itself
can have singularities on it.  We will primarily restrict our attention
to the case in which this additional complication does not arise,
and assume that the 4-cycle is actually a complex manifold.

Under this assumption, the del Pezzo surfaces are easy to classify:
each such surface is either ${\bf P}^1\times{\bf P}^1$, or the blowup of
${\bf P}^2$ at $d$ general points, $0\le d\le8$.  If $S$ is such a surface,
then $H^{1,1}(S)$ has rank $d+1$ (rank 2 in the case of
${\bf P}^1\times{\bf P}^1$),
with generators consisting of a line from ${\bf P}^2$, and the
exceptional divisors of the blowups (or the two ${\bf P}^1$'s in the case
of ${\bf P}^1\times{\bf P}^1$).
When $S$ appears on a Calabi--Yau threefold $X$, the image of the
natural map $H^{1,1}(X)\to H^{1,1}(S)$ has some rank $k$ with
$1\le k\le d+1$, which is an important invariant of the situation.

Each such surface $S$ on $X$ will lead to a potential extremal transition,
provided that $S$ is contracted to a point when some wall of the K\"ahler
cone is approached.  The structure of this extremal transition appears
to be this:\footnote{These statements have not been fully verified in the
mathematics literature, but supporting evidence for the picture we
present here can be found in \cite{namikawa,gross}.  Experts please
note that when $k>1$ we are treating non-$\Q$-factorial singularities.}
the singular point can be smoothed through complex structure deformations
to a new Calabi--Yau threefold $X'$
leading to a new branch of the moduli space with Hodge numbers
$$h^{1,1}(X')=h^{1,1}(X)-k$$
$$h^{2,1}(X')=h^{2,1}(X)+c_d-k,$$
where $c_d$ is an invariant of the singularity type which we will discuss
further below.

The surfaces $S$ possess some rather remarkable geometric
structures \cite{demazure,manin}.
Assume for now that $d\ge 3$ and $S\ne{\bf P}^1\times{\bf P}^1$, and let
$\ell$, $e_1$, \dots, $e_d$ denote the generators of $H^{1,1}(S)$.
Then $c_1(S)=3\ell-\sum e_i$, and if we consider the lattice
$$\Lambda_d:=\{x\in H^{1,1}(S,\Z) \ |\ x\cdot c_1(S)=0\},$$
then $\Lambda_d$ is naturally isomorphic to the root lattice of
the root system $E_{d}$, with generators $\ell-e_1-e_2-e_3$,
$e_i-e_{i+1}$.  We thus get a natural action of the Weyl group
$W(E_{d})$ on $H^{1,1}(S)$.  One interesting feature of this Weyl
group action is that it permutes the classes $\lambda\in H^{1,1}(S,\Z)$
which satisfy $\lambda\cdot\lambda=-c_1(S)\cdot\lambda=-1$ (these
can be thought of as the ``lines'' on $S$).  The number of these
``lines'' is $6, 10, 16, 27, 56, 240$ for $d=3, 4, 5, 6, 7, 8$,
respectively.  The number $c_d$ which measures the change in Hodge
numbers for $S$ embedded on a Calabi--Yau threefold is given by
the dual Coxeter number of $E_d$.

The singularities which lead to these del Pezzo surfaces upon a
crepant blowup can also be described quite explicitly \cite{reid} (at least
for large values of $d$).  They behave as follows:

(i) For $d=8$ the singularity is locally a hypersurface in ${\bf C}^4$ whose
equation has leading terms $y^2=x^3+e_2(s,t)x^2
+f_4(s,t)x+g_6(s,t)$, where $e$, $f$ and $g$ are homogeneous polynomials
of the indicated degrees.

(ii) For $d=7$ the singularity is locally a hypersurface in ${\bf C}^4$ whose
equation has leading terms $y^2=f_4(x,s,t)$
where $f_4$ is a homogeneous polynomial defining a nonsingular curve
in ${\bf P}^2$.

(iii) For $d=6$ the singularity is locally a hypersurface in ${\bf C}^4$ whose
equation has all of its leading terms of degree three,
defining a nonsingular surface of degree 3 in ${\bf P}^3$.

(iv) For $d=5$ the singularity can be locally represented in ${\bf C}^5$
as the intersection of two hypersurfaces, both of whose leading terms
are nondegenerate homogeneous quadratic polynomials.

(v) For $d\le4$, the singularity cannot be locally represented as
a transverse intersection of hypersurfaces in some space.

\subsection{Extremal transitions for elliptically fibered Calabi--Yau
threefolds}

We now consider which extremal transitions will occur for elliptically
fibered Calabi--Yau threefolds.  The K\"ahler parameters which we
wish to vary in order to approach a transition point are the K\"ahler
parameters coming from the base of the Calabi--Yau manifold.
We wish to shrink some curve on the base to zero size, and then ask
whether there is an extremal transition to another elliptically
fibered Calabi--Yau threefold.  In order for such a transition to
exist, we must be able to smooth the base through complex deformations,
in a way compatible with the existence of an elliptic fibration.
The only curves on the base for which such a transition is possible
are rational curves with self-intersection $-n$, for $n=1, 2, 4$.
This is because the local functions $f$ and $g$ used in the Weierstrass
equation must be sections of line bundles $-4K_B$ and $-6K_B$ on
the singular base $B$ if the smoothing is to make sense,
but in order for those multiples of $K_B$
to be line bundles, we must have $4(n-2)/n$ and $6(n-2)/n$ both being
integers.\footnote{Before shrinking this curve down, we have
$-K_B\cdot C=n-2$.  Thus, we need for $m(n-2)/n$ to be an integer in order
that
$-mK_B - \frac{m(n-2)}n\,C$ will define
a  divisor which is orthogonal to $C$ and so can come from
a line bundle after blowing down.}  This is only possible if $n$ divides $4$.

Note that in the case of elliptic Calabi--Yau manifolds over ${\bf F}_n$
the strong coupling transition point gets mapped to where a rational
curve (denoted by $D_v$ in \cite{mv}) shrinks
to zero size and that curve has self-intersection $-n$.  The fact that for the
cases of $1,4$ there could be transition has been anticipated in \cite{sw}
based on anomaly considerations.  It is satisfying to see that
the mathematical conditions for transitions are captured by the
anomaly considerations.  The case $n=2$ is special and as  has been
argued in \cite{mv} it corresponds upon further compactification to
4 dimensions to $SU(2)$ gauge symmetry with an adjoint matter.

In order to determine the structure of these possible extremal transitions,
we need to study an elliptic fibration defined on a neighborhood of
a rational curve $C$ of self-intersection $-n$.  In fact, this local study
can be globalized, since we have available the surface ${\bf F}_n$
with elliptic fibrations over it which are sufficiently general to
reproduce all possible local behaviors in a neighborhood of the
curve of negative self-intersection.  So we shall study the extremal
transitions for those more global models, i.e., taking $C=D_v$
in the ${\bf F}_n$ model.

The $n=2$ case was studied in detail in \cite{mv}; let us
review the main points.  The curve $C$ satisfies $K_B\cdot C=0$,
and so the locus where the fibers become singular does not meet $C$.
It follows
that the elliptic fibration over $C$ is a product $C\times T^2$
with a fixed complex structure on $T^2$.  In particular, we can
blow down $C\times T^2$ along the $C$ direction, leaving a curve
of singularities of type $A_1$, and so (in the 4-dimensional theory)
an $SU(2)$ enhanced gauge symmetry.  The curve itself has genus one,
so from the analysis of \cite{KM,KMP} we expect a matter content
of one adjoint hypermultiplet.  At generic moduli the $SU(2)$ is
broken, and the base surface must be deformed.  In the global
model over ${\bf F}_2$, this deformation
corresponds to a ``non-polynomial deformation''
of the $WP_{1,1,2,8,12}^4$ model, in which the contracted ${\bf F}_2$
deforms to the smooth surface ${\bf F}_0={\bf P}^1\times{\bf P}^1$.

The $n=4$ case bears a number of similarities to the previous one.  In order
to satisfy the adjunction formula, $C$ must appear in the divisor
for $-K_B$ with coefficient $1/2$.  It follows that along
the curve $C$, we must have an $SO(8)$ enhanced gauge symmetry.
The Weierstrass fibration will have as its fiber a rational curve
with a cusp, and the curve of cusps is a singular curve of the total
space of type $D_4$.  We can blow this down in the $C$ direction;
the result will be a curve of $A_1$ singularities which is a rational
curve with a cusp.  There is a {\it new}\/ $SU(2)$ gauge symmetry
associated with this, in addition to the $SO(8)$ which is present
for all values of moduli.
At the cusp is hiding a rather complicated singularity, since the entire
previous curve of $D_4$ singularities has mapped to this single point.

To see that the resulting space can be smoothed, we consider the global
model.  After contracting $C$ to a point, the base surface can be
represented by $WP_{1,1,4}^2$ and the singular Calabi--Yau is a
hypersurface
in $WP_{1,1,4,12,18}^4$.  There is a 2-1 cover of this base
surface by $WP_{1,1,2}^2$, which is ${\bf F}_2$ with {\it its}\/
$D_v$ contracted
to a point.  That is, we can realize $WP_{1,1,4}^2$ in the form
$WP_{1,1,2}^2/{\bf Z}_2$.  The fixed points of the ${\bf Z}_2$ consist
of the singular point of $WP_{1,1,2}^2$, and another curve which does
not pass through that point.  The ${\bf Z}_2$-action can be lifted to
the Weierstrass fibration, where it acts as $-1$ on the elliptic fibers.
In terms of the homogeneous coordinates $[s,t,u,x,y,z]$ on
$WP_{1,1,2,8,12}^4$,
the ${\bf Z}_2$ acts as $-1$ on $u$ and $y$, and fixes the other variables.
We can explicitly map the quotient to $WP_{1,1,4,12,18}^4$ by
$$[s,t,u,x,y]\mapsto[s,t,u^2,u^2x,u^3y],$$
and note that a Calabi--Yau in the first space is mapped to a Calabi--Yau
in the second.
If we smooth $WP_{1,1,2}^2$ to ${\bf P}^1\times{\bf P}^1$,
the ${\bf Z}_2$ action follows the smoothing and acts to swap the
two ${\bf P}^1$ factors in ${\bf P}^1\times{\bf P}^1$.  The
quotient by that is ${\bf P}^2$; we see in this way that the ${\bf F}_4$
models with $D_v$ contracted smooth to the ${\bf P}^2$ model.

The extremal transition, blowing down to a curve of $A_1$'s with a
cusp $P$, is illustrated in figure 2.  This is similar to, but apparently
different from, an extremal transition discussed by
Aspinwall \cite{AspinExtreme} in a similar context.  Aspinwall was
interested in preserving a K3 fibration, and so used a singular model
of the elliptic fibration in which 4 disjoint surfaces over $C$ had
been blown down to 4 curves of $A_1$ singularities (leading to a
gauge symmetry enhancement of $SU(2)^4$) rather than the 4 surfaces
we blow down in the Weierstrass model which lead to a single curve
of $D_4$ singularities.

\iffigs
\begin{figure}
  \centerline{\epsfxsize=11cm\epsfbox{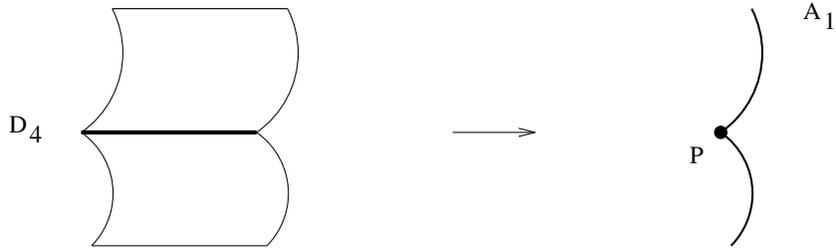}}
  \caption{The extremal contraction with $n=4$.}
  \label{fig2}
\end{figure}
\fi

We turn now to the extremal transitions in the case $n=1$.
The elliptic fibration over $C$ defines a rational elliptic surface,
generically with 12 singular fibers.  We also have a distinguished
section on that surface, the intersection of the global section of
the Weierstrass fibration with the surface.  Unlike the previous cases,
this surface cannot be directly contracted to a curve, since the
fiber type of the elliptic fibration varies.  In order to learn how
this transition works, we study the global model of the extremal transition
from ${\bf F}_1$ to ${\bf P}^2$.
Following the discussion in \cite{mv}, the (3,243)
Calabi--Yau threefolds with an elliptic fibration over ${\bf F}_1$
can be described as hypersurfaces in an ambient space determined by
the following data:  there is a space ${\bf C}^7$ with `homogeneous'
coordinates $s,t,u,v,x,y,z$ from which we remove the loci
$\{s=t=0\}$, $\{u=v=0\}$, $\{x=y=z=0\}$ and then form
quotient by the $({\bf C}^*)^3$
action specified by the following table.

\begin{center}
\begin{tabular}{|c|ccccccc|}\hline
&$s$&$t$&$u$&$v$&$x$&$y$&$z$\\ \hline
$\lambda$&$1$&$1$&$1$&$0$&$6$&$9$&$0$\\ 
$\mu$&$0$&$0$&$1$&$1$&$4$&$6$&$0$\\ 
$\nu$&$0$&$0$&$0$&$0$&$2$&$3$&$1$\\ \hline
\end{tabular}
\end{center}
(The entries in the table denote exponents for the action of the ${\bf
C}^*$'s
on the homogeneous variables.)  This ambient space is in fact a toric
variety, and we can use the methods of toric geometry (see \cite{agm} for
a review)
to study the K\"ahler moduli.  The columns in the table represent
elements of $H^{1,1}$; in a more convenient basis, the action would
be described by the following table.
\begin{center}
\begin{tabular}{|c|ccccccc|}\hline
&$s$&$t$&$u$&$v$&$x$&$y$&$z$\\ \hline
$\lambda\mu^{-1}\nu^{-1}$&$1$&$1$&$0$&$-1$&$0$&$0$&$-1$\\
$\mu\nu^{-2}$&$0$&$0$&$1$&$1$&$0$&$0$&$-2$\\
$\nu$&$0$&$0$&$0$&$0$&$2$&$3$&$1$\\ \hline
\end{tabular}
\end{center}
The columns of the matrix again represent generators of $H^{1,1}$;
this basis has been chosen so that the K\"ahler cone is spanned by $(1,0,0)$,
$(0,1,0)$ and $(0,0,1)$.

Our extremal transition should be described by shrinking the K\"ahler
parameter corresponding to the curve of self-intersection $-1$
in the base, and this corresponds in the toric description to
approaching the wall of the K\"ahler cone spanned by the second and third
coordinates.  When we approach that wall, however, we find that a
2-cycle shrinks to zero size rather than a 4-cycle.  In fact,
crossing that wall produces a ``flop''
of the Calabi--Yau space to another Calabi--Yau threefold, whose
K\"ahler cone is spanned by $(0,1,0)$, $(0,0,1)$, and $(-1,1,1)$.
(This flop is clearly signaled by the first row of our matrix of charges,
which has as it's nonzero entries $1$, $1$, $-1$ and $-1$.)
After we make this flop, the resulting space maps to ${\bf P}^2$ rather
than to ${\bf F}_1$, and the general fiber of the map is still a $T^2$,
but the fiber over one point in ${\bf P}^2$ contains an entire
4-cycle.  To complete the transition, then, we must also shrink
this 4-cycle to zero size, and that is accomplished in the toric
setup by proceeding to the wall of the new K\"ahler cone which is
spanned by $(0,1,0)$ and $(-1,1,1)$.  Along that wall, the 4-cycle
has shrunk to zero size, and we can map the resulting threefold into
${\bf P}^{1,1,1,6,9}$ where its singularity can be smoothed by a
complex structure deformation.  Doing so goes to a different branch
of the 4-dimensional moduli space.

It is worthwhile noting what lies on the other side of that second
wall: we enter
a somewhat non-geometric phase in the moduli space of the 4-dimensional
theories, one in which one of the K\"ahler parameters has been replaced
by some field localized at the singular point.  The cone in $H^{1,1}$
associated with this new phase is spanned by $(0,1,0)$, $(-1,1,1)$
and $(-1,1,0)$.  All three cones---which constitute a part of the
enlarged K\"ahler moduli space for this---are illustrated in figure 3,
with the F-theory boundary shaded.

\iffigs
\begin{figure}
  \centerline{\epsfxsize=11cm\epsfbox{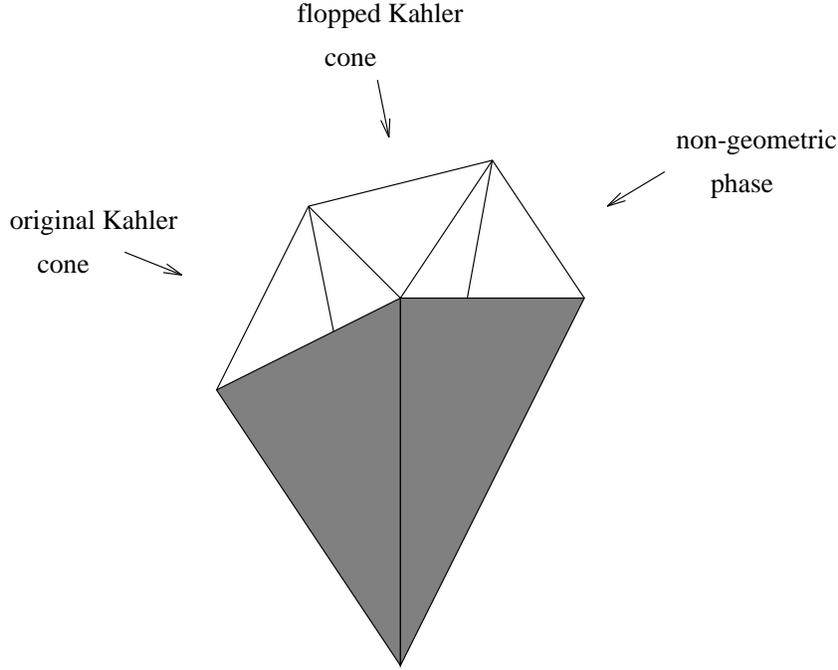}}
  \caption{The F-theory moduli space on the boundary of the
enlarged K\"ahler moduli space.}
  \label{fig3}
\end{figure}
\fi

Note that in the limit where the fiber shrinks to zero (so that
we get the F-theory moduli space)
the second phase region lies on the boundary of the first.  The
third phase region shows up in the F-theory moduli space as the
region where the area of $D_v$ has been formally extended to negative
values, but as this is not a geometric region, it is not clear that
there is actually an associated 6-dimensional theory.

We can describe this transition more abstractly as follows.  Initially,
lying over the curve of self-intersection $-1$ was a rational elliptic
surface with a section, and such a surface must be the blowup of
${\bf P}^2$ in 9 points.  In our transition, we first performed a flop
on one of the `lines' on this surface, leaving a divisor which is
the blowup of ${\bf P}^2$ in 8 points, i.e., a del Pezzo surface with
$d=8$.  We then performed an extremal transition on that del Pezzo
surface, leaving a singular point $P$.  (This is illustrated in figure 4.)
The structure of the transition is plainly visible in
the behavior of the equations of the threefold during this process.
The original equation took the form
$$y^2=x^3+\sum_{k=-4}^{4} xz^{4-k}f_{8-k}(z') +\sum_{k=-6}^{6}
z^{6-k}g_{12-k}(z')$$
and has its terms constrained by the need to embed into the toric
variety that fibers over ${\bf F}_1$.
When we embed it into ${\bf P}^{1,1,1,6,9}$,
however, we find that additional terms are permitted as new complex
deformations:
$$\sum_{k=5}^8xf_{8-k}(z') + \sum_{k=6}^{12}g_{12-k}(z').$$
In fact, if $z''$ is the third homogeneous coordinate on ${\bf P}^2$,
we see that before adding in these new terms,
there is a singularity at $z'=z''=0$, where the leading
terms are
$$y^2=x^3+xF_4(z',z'')+G_6(z',z''),$$
where $ F_4(z',z'')$ and $G_6(z',z'')$ are homogeneous polynomials in
$z'$ and $z''$.  This is recognized immediately as the `canonical
singularity with crepant blowup' with $d=8$.

\iffigs
\begin{figure}
  \centerline{\epsfxsize=13cm\epsfbox{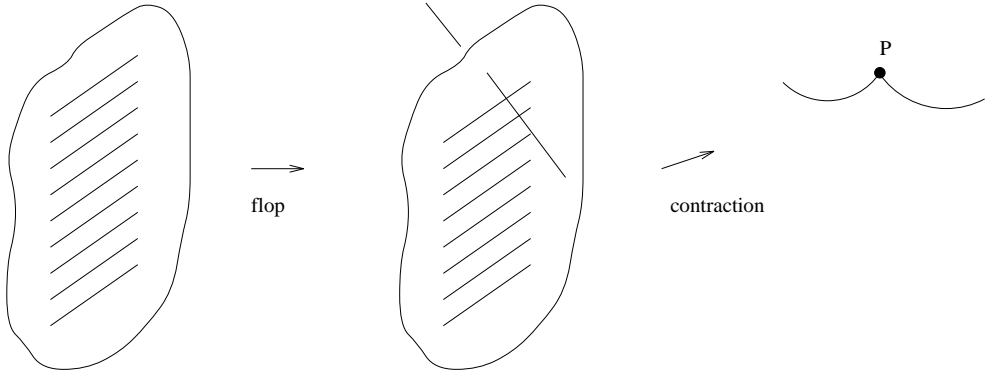}}
  \caption{The flop and extremal contraction with $n=1$.}
  \label{fig4}
\end{figure}
\fi

There is a natural generalization of this transition: instead of first
flopping along a single section, we could flop along $k$ sections,
leaving a del Pezzo surface with $d=9-k$.  The extremal transition
along that del Pezzo surface could then be done; as in the earlier case,
the F-theory moduli space sees only the final result
and not the intermediate steps.  The new model
produced in this way would have a larger number of sections of its
elliptic fibration, and a smaller value of $h^{2,1}$.

\subsection{Applications}

We can now also discuss transitions among the elliptic Calabi--Yau
manifolds over ${\bf F}_n$.    The basic set up for this
has been suggested in \cite{sw} based on M-theory considerations:
We go to a region in hypermultiplet moduli where we get an $E_8$
instanton of zero size.  This is the region where we can blow up
and obtain one additional tensor multiplet whose scalar controls
the value of the K\"ahler class of the blow up.  The basic transition
of this type is of the $E_8$ type corresponding to the $d=8$ transition
discussed in detail above; in other words the basic setup of the
transition is exactly the same one responsible for going from ${\bf F}_1$ to
${\bf P}^2$ (or backwards).  This can be used to do two things:  We can
go from ${\bf F}_n$ to ${\bf F}_{n\pm 1}$ by blow ups and blow downs, or we can
just blow up as many of the $E_8$ instantons as we please.  Of course
when we are left with less than instanton number 4 in either $E_8$ there
will not exist any hypermultiplet moduli corresponding to giving a size
to the instantons.  In that case the instantons are necessarily all of zero
size; this should be viewed as a boundary limit where all the
instantons are blown up to tensor multiplets with the K\"ahler
classes of some of the tensor multiplets tuned to correspond to zero size
$E_8$ instantons.
We will now discuss in detail how the ${\bf F}_n$ to ${\bf F}_{n\pm 1}$
transition works.

Recall that in \eqref{genexbl} we blew up $z=z'=0$ in ${\bf F}_n$ by
introducing a new variable $u$ and a ${\bf C}^*$, and rewriting the
equation
as
\begin{equation}\label{genexbis}
y^2=x^3+\sum_{k=-4}^{4}\sum_{\ell=0}^{8-nk} f_{k\ell}\,z'{}^\ell z^{4-k}
u^{\ell-k}x
+\sum_{k=-6}^{6}\sum_{\ell=0}^{12-nk} g_{k\ell}\,
z'{}^\ell z^{6-k}u^{\ell-k}.
\end{equation}
To return to ${\bf F}_n$, we would simply set $u=1$; if instead we wish
to map to ${\bf F}_{n+1}$ we can simply set $z'=1$.  This produces
the equation for a Calabi--Yau fibering over that space.

In the process of doing this, we used one of the extremal transitions
with $n=1$.  After passing to ${\bf F}_{n+1}$, it may be possible
to further deform the equation by adding new terms; this can affect
the generic gauge symmetry of the model.

As another application, we can apply $n=1$ transitions to the $(19,19)$
model.\footnote{These transitions were also analyzed in some detail by Borcea
\cite{borcea}
from a slightly different perspective (see also \cite{schoen}).}  
We can blow down $(-1)$-curves
successively until we get to the ${\bf P}^2$ model.  At each stage, using
the $d=8$ transitions, the Euler characteristic of the Calabi--Yau manifold
changes by 60.

Finally, the transitions with $n=4$ can be used to relate the various
Voisin--Borcea models.  The rational curves occurring on those models
when $k>0$ are curves of self-intersection $-4$, and if we make an extremal
transition we alter the invariants as $k\mapsto k-1$, $g\mapsto g$, which
is achieved via $(r,a)\mapsto(r-1,a+1)$.  This links together large numbers
of the transitions; further links can be made when $k=0$ by using the
$n=1$ transitions.

\subsection{Physical aspects of transitions}

We would like to discuss some physical aspects of transitions
and singularities encountered in the moduli space of the theory
discussed in the mathematical setup in the previous section.
The main case of interest is the singularities and potential
transitions in the moduli space of tensor multiplets, which
have a scalar.  As discussed before the scalars in the tensor
multiplets correspond to K\"ahler classes of the base manifold
in the F-theory setup.  Singularities are expected when
the K\"ahler classes are tuned such that some 2-cycle in the base has vanishing
area.  For example, as noted in \cite{mv}, in the case of
F-duals of perturbative heterotic vacua one finds elliptic fibrations
over
${\bf F}_n$.  If $n\not=0$ it was shown in \cite{mv} that
there is a 2-cycle, denoted by
$D_v$ in \cite{mv} whose area goes as $k_b-{n\over 2}k_f$,
where $k_{b,f}$ denote the K\"ahler classes of the base/fiber
of ${\bf F}_n$.  The scalar in the tensor multiplet controls $k_b/k_f$
and so when that reaches $n/2$ we have a vanishing 2-cycle.  This
in particular was the F-theory realization of the
strong coupling puzzle raised in \cite{dmw}.   In particular
the kinetic term of the gauge theory behaves as
$$\int (k_b-{n\over 2}k_f)F^2$$
How do we understand this in the F-theory set up? This
is essentially straightforward because the 2-cycle in question
is part of the 7-brane world volume which is responsible
for the gauge symmetry.  In particular the 7-brane worldvolume
is given by $D_v\times R^6$, thus the effective coupling
constant for the gauge field in $R^6$ has a term proportional
to the volume $D_v$ which thus explains the above singularity.
The existence of this singularity is also marked by the
existence of tensionless strings \cite{sw,dufet}.
{}From the viewpoint of a Lagrangian formulation
these are simply the instantons
of the gauge group which have zero action in the limit of
infinitely strong coupling, thus leading to tensionless strings.
{}From the viewpoint of F-theory these tensionless strings are the
threebranes wrapped around the vanishing cycle $D_v$.  Note that
this is consistent with the instanton interpretation because
a 3-brane wrapped partially around $D_v$ and living as a string
in $R^6$ is itself
inside a 7-brane  and it is well known that one can view
instantons living on D-branes as an equivalent description
for the 4 lower dimensional D-branes sitting inside them
\cite{doug,dins}.

The main question is what happens after we hit such a wall
where we have tensionless strings.  In principle three things
can happen: 1-Nothing, we are stuck at that point.  2-
Continue formally to larger values of the K\"ahler parameter
which formally would have yielded negative kinetic term
for the gauge field.  3-Make a transition to a new branch
of $N=1$ theories in 6 dimensions.
For case (1) we have nothing to say and this case may in fact be the
generic case.
For case (2) we note
that this is what one may expect in a lower dimensional theory
such as compactifications down to 5 or 4 dimensions, for which
the moduli of real K\"ahler moduli or complex K\"ahler moduli
can be continued beyond the K\"ahler walls \cite{agm,phases}.
  However the question is whether they make sense
in 6-dimensional terms.   In the 4 or 5 dimensional
cases typically the transition is to non-geometric
phases, so it seems unlikely that in such regimes
the 6 dimensional version can be defined. In particular,
in general there may be no notion of taking the K\"ahler
class of the elliptic fiber to be small.
If the 4d transition is to phases which are geometric enough and moreover
admit an elliptic fibration then one could in principle take
the small K\"ahler class limit and recover a 6d transition.  Otherwise
it is difficult to justify the existence of 6-dimensional transitions.

As for point (3) above that is the most interesting
aspect of these singularities.  In the previous subsection
we have given a mathematical analysis of the transition
types available.  In particular we found that there are only
two basic types, corresponding to vanishing spheres with
self-intersection $-n$ for $n=1,4$ (as discussed before the
case $n=2$ does not lead to a transition). These two
transitions were qualitatively very different because in
the case of $n=4$  in 4 dimensions we get a curve of singularities
on Calabi--Yau which signal an enhanced gauge symmetry point including an extra
 $SU(2)$
with possibly some matter content.  In 4d terms this can in principle
be some kind of Higgs mechanism, though it has to be studied further.
In the $n=1$ case we have a very
different type of transition which is bound to contain new physics
as we will discuss now.  Recall that the $n=1$ case
has many subcases according to the geometry of the vanishing
4-cycle $S$ in the Calabi--Yau manifold.
We discussed the classification of the simplest types
of them in the previous section in terms of del Pezzo surfaces corresponding
to $E_d$ groups.  In the previous section we have used the $E_8$ type
del Pezzo surface to go between various $F_n$'s.
 In fact it has been argued
in \cite{sw,han}\ that such transitions are also natural from the viewpoint
of M-theory:  In particular a 5-brane approaching the boundary
of the 11-th dimension will give rise to a tensionless string
(the membrane stretched from the 5-brane to the boundary).  Moreover
when it is at the boundary it can be viewed as an $E_8$ instanton
of zero size.  We can then give the instanton a finite size, thus
making a transition to a model with one less tensor multiplet and
(generically) 29 more hypermultiplets.  Moreover it was noted in \cite{sw}
that the same process can be described by blow up/down in the
F-theory set up.
Both the M-theory and the F-theory setup thus suggest that
such transitions are natural.  But unfortunately neither one
really explains the physics of the transition.
  Let us try to develop the physics
of transition as seen from the 4d perspective.
 In 4d the fact that a 4-manifold $S$ has shrunk
to a point, means that we have  the 2-cycles in $S$ shrinking
as well as the 4-cycle $S$ itself shrunk to a point.
It seems we have massless states corresponding to all the vanishing
2-cycles (which are realized as holomorphic curves) as well as the 4-cycle
itself. Let $C$ be a 2-cycle realized by a holomorphic curve.
Consider the intersection number $C\cdot S=C\cdot c_1(S)$,
which is frequently non-zero.
Intersecting vanishing
2- and 4-cycles signifies the appearance of massless electric
and magnetic hypermultiplets.  These are the analogues (
and perhaps extensions) of Argyres--Douglas \cite{ado}
points of $N=2$ theories
in string theory \cite{witco}.
  Let us assign
magnetic $U(1)$ charge $1$ to the 4-cycle $S$.
If $C\cdot S=k$ we then see that $C$ has electric charge $k$
under the same $U(1)$.
Unfortunately we do not have a
Lagrangian description of such points and thus the description
of the transitions in this language is an interesting open problem.
However there seem to be a number of additional hints which may
shed light on the nature of the
 transitions we are encountering.  As discussed above, among the
states with $k=0$ (i.e., 2-cycles with $C\cdot S=0$)
 which are thus not charged under the $U(1)$, the
root system of $E_d$ group naturally appears (as lines).
Moreover if we consider a 2-cycle of fixed genus $g$ and electric
charge (i.e., intersection number with $S$ equal to)
 $k$ under $U(1)$, the Weyl group of $E_d$ acts on them.  Thus
they can be taken to form a representation of $E_d\times U(1)$
(if we also include the zero-weight states),
where the $E_d$ representation is fixed by the Weyl group action
and the $U(1)$ charge is electric equal to $k$.  For example
as discussed in the previous section the $g=0,k=1$ case
form well known representations
of $E_d$ (the fundamental representation for $d=8,7,6$, the spinor
for $d=5$ the anti-symmetric tensor representation for $d=4$ and
$(3,2)$ representation for $E_3=SU(3)\times SU(2)$).  These facts suggest
that at these transition points we
have an enhanced $E_d$ gauge symmetry together with some electric/magnetic
quantum numbers of a $U(1)$.  Note also that the representations
are expected to grow larger as we consider arbitrary $g$ and $k$.
We thus seem to get infinitely many massless modes which realize
ever increasing representations of $E_d$.  This sounds very
similar to the story of how we get a huge number of higher genus 2-branes
inside a $K3$ \cite{bbsv}.
The story is similar here in that
(ignoring the bundle choice of the cycle) the expected dimension
of the family grows with genus (as $k+g-1$) and we thus expect
a huge number of 2-brane states.
   However it is different in that
here apparently all of them are becoming massless (as they
are vanishing cycles) as we approach
the transition point.  Given the exponential growth expected
in such states and the fact that we have a natural $E_d$ representation
for all of them we conjecture that these states organize themselves
into affine Kac--Moody algebras of the $E_d$ group (where
the grading is related to $k,g$).\footnote{The Weyl group of this affine
Kac--Moody algebra occurs naturally in the study of del Pezzo surfaces
\cite{looijenga,pinkham}.}   We are currently
investigating the validity of this conjecture.  Note that this conjecture
is also in line with the idea that the 6 dimensional version
of the transition is through tensionless strings. Thus if the
tensionless string carries an $E_d$ current algebra
we would naturally get the above prediction where
essentially the entire affine $E_d$ representation
has seemingly collapsed to zero energy because the string has no tension.

Note that in the previous sections we showed how the $E_8$ case
of these transitions allows one to go between ${\bf F}_n$'s and
corresponds to points where an $E_8$ instanton has shrunk to
zero size.  In this sense one can expect that we should
at least locally restore the $E_8$ symmetry that was broken
by a finite size instanton.  Here what we seem to be finding
is that this seems to be the case with two subtleties that
upon compactification to 4d we
have not only infinitely many
 electric representations of the left-over $E_8$
but also some of them have relatively
non-local charge with respect to the 4-cycle $S$.
Similarly it is thus natural to conjecture that the other
$E_d$ cases are also $F$-duals of $E_d$ instantons
shrinking to zero size leaving us with a superconformal
point of an $N=2$ theory with massless electric and magnetic states.
It is important to study the physical meaning of transitions through
such points.  This is a case where at the present the
mathematics of the transition is better understood than the physics of it.
Hopefully we have provided some hints about the physics of it which
  will guide us in developing an effective physical formulation
of transitions in $N=2$ superconformal field theories.  Note
that even though we may associate the $E_d$ cases with an instanton
of $E_d$ of zero size, once we think of it as a heterotic fivebrane
(corresponding to singularities of (0,4) superconformal theories)
it has a life of its own and may appear in transitions where
formally no gauge group exists.  We have seen examples of this
in the context of blow downs of the (19,19) model.

We would like to thank M. Bershadsky, B. Greene, M. Gross, A. Strominger
and E. Witten for valuable discussions.
 D.R.M.
would also like to thank the Harvard Physics Department for hospitality
during the preparation of this paper.
The research of D.R.M. is supported in part by NSF grant DMS-9401447,
and that of C.V. is supported in part by NSF grant PHY-92-18167.

\end{document}